%% file: Journal_TCOMMS.tex
\documentclass[12pt,onecolumn,final]{IEEEtran}
\usepackage[nomarkers,nolists,noheads]{endfloat}
\usepackage{cite}
\ifCLASSINFOpdf
  \usepackage[pdftex]{graphicx}
 
  \graphicspath{{../Figures/}}
  \DeclareGraphicsExtensions{.eps}
\else
  
  \usepackage[dvips]{graphicx}
  
\fi
\usepackage[cmex10]{amsmath}
\usepackage{algorithmic}
\usepackage[tight,footnotesize]{subfigure}
\usepackage{amssymb}

\begin{document}
\title{Secrecy Sum-Rates for Multi-User MIMO \\ Regularized Channel Inversion Precoding}

\author{\normalsize\authorblockN{{Giovanni~Geraci$^{1,3}$,~Malcolm~Egan$^{2,3}$,~Jinhong~Yuan$^1$,~Adeel~Razi$^{1,3}$~and~Iain~B.~Collings$^3$}}\\
\small\authorblockA{$^1$School of Electrical Engineering \& Telecommunications,
The University of New South Wales, \textsc{Australia} }\\
\authorblockA{$^2$School of Electrical and Information Engineering, The University of Sydney, NSW, \textsc{Australia}}\\
\authorblockA{$^3$Wireless and Networking Technologies Laboratory, CSIRO ICT Centre, Sydney, \textsc{Australia}}
}
\maketitle

\input{Abstract}
\IEEEpeerreviewmaketitle
\thispagestyle{empty}

\newpage
\input{Section1}

\input{Section2}

\input{Section3}

\input{Section4}

\input{Section5}

\input{Section6}

\input{Section7}
\ifCLASSOPTIONcaptionsoff
  \newpage
\fi

\bibliographystyle{IEEEtran}

\bibliography{IEEEabrv,Bib_Giovanni}
\end{document}

%% file: Abstract.tex
\begin{abstract}
In this paper, we propose a linear precoder for the downlink of a multi-user MIMO system with multiple users that potentially act as eavesdroppers. The proposed precoder is based on regularized channel inversion (RCI) with a regularization parameter  $\alpha$ and power allocation vector chosen in such a way that the achievable secrecy sum-rate is maximized. We consider the worst-case scenario for the multi-user MIMO system, where the transmitter assumes users cooperate to eavesdrop on other users. We derive the achievable secrecy sum-rate and obtain the closed-form expression for the optimal regularization parameter $\alpha_{\mathrm{LS}}$ of the precoder using large-system analysis. We show that the RCI precoder with $\alpha_{\mathrm{LS}}$ outperforms several other linear precoding schemes, and it achieves a secrecy sum-rate that has same scaling factor as the sum-rate achieved by the optimum RCI precoder without secrecy requirements. We propose a power allocation algorithm to maximize the secrecy sum-rate for fixed $\alpha$. We then extend our algorithm to maximize the secrecy sum-rate by jointly optimizing $\alpha$ and the power allocation vector. The jointly optimized precoder outperforms RCI with $\alpha_{\mathrm{LS}}$ and equal power allocation by up to $20$ percent at practical values of the signal-to-noise ratio and for $4$ users and $4$ transmit antennas.
\end{abstract}

%% file: Section1.tex
\section{Introduction}
\setcounter{page}{1}
In current practical multi-user MIMO systems such as LTE and 802.11n, securing transmitted data from nearby eavesdroppers is critical. In these systems, security is achieved using potentially vulnerable network layer cryptography techniques. The vulnerability is due to a reliance on the limited resources of the eavesdropper and on the unproven computational complexity of inverting the encryption algorithms \cite{Mukherjee10Survey}. To enhance the protection of transmitted data and achieve perfect secrecy, methods exploiting the channel, known as physical layer security, have been proposed.

Physical layer security techniques were proposed to protect the data from eavesdroppers for several network topologies in \cite{Wyner75,Csiszar78,LiCISS07,KhistiISIT07,Khisti10I,
Mukherjee11Robust,Goel08,Zhou10,Lin11,Zhou11,Khisti08,Ekrem11,Liu10}. In \cite{Wyner75,Csiszar78}, a three-terminal network consisting of a transmitter, an intended user and an eavesdropper, known as the wiretap channel, was considered. The authors derived the secrecy capacity, where the message is transmitted reliably to the intended user while the rate of information leakage to the eavesdropper vanishes asymptotically with the code length. The secrecy capacity of the wiretap channel was derived for the MIMO case in \cite{LiCISS07,KhistiISIT07,Khisti10I
} when all terminals had full channel state information. It was shown in \cite{Mukherjee11Robust,Goel08,Zhou10,Lin11,Zhou11} that the transmission of artificial noise, as well as adaptive encoding, is an effective method to reduce the eavesdropper's signal-to-noise ratio when the eavesdropper's channel is not known by the transmitter. Recently, physical layer security was also extended to multiuser networks where the eavesdropper is not an intended user \cite{Khisti08,Ekrem11} and to two-user networks where the intended users are also eavesdroppers \cite{Liu10}. The secrecy capacity region for multi-user networks where any number of intended users are potentially eavesdropping remains an open problem. Moreover, the achievable secrecy rates of such multi-user networks with practical transmission schemes are also unknown.

\setcounter{page}{1}

Suboptimal precoding schemes have proven to be practical and effective in controlling interuser interference for the downlink of multi-user MIMO networks \cite{YooJSAC06,Spencer04,Peel05,Joham05,Jang08,Sun10,HochwaldVP05,Razi10,Schmidt05}. While the sum-capacity of multi-user MIMO networks without eavesdroppers is achieved using dirty paper coding \cite{CaireZFDPC03}, it requires high-complexity coding
schemes \cite{Spencer04Magazine}. Linear precoding schemes were proposed as a low-complexity alternative for multi-user MIMO downlink implementations \cite{Li10a}. A popular and practical linear precoding scheme to control interuser interference is channel inversion (CI) precoding, sometimes known as zero forcing precoding \cite{YooJSAC06,Spencer04}. To increase the sum-rate performance of the CI precoder, the regularized channel inversion (RCI) precoder was proposed to tradeoff the interuser interference and the desired signal through a regularization parameter \cite{Peel05}. 
Linear precoding schemes were also proposed to achieve secrecy in single-user MIMO networks \cite{LiCISS07,KhistiISIT07,Khisti10I,Mukherjee11Robust}.

In \cite{GeraciISWCS11}, the use of linear precoding was proposed to achieve physical layer security in a multi-user MIMO system. For RCI precoding, the authors obtained an achievable secrecy sum-rate as a function of the singular values of the channel. A bound on the optimal regularization parameter of the precoder was also given in the large-system regime.

In this paper, we consider the multi-user MIMO downlink with multiple single-antenna users that cooperate and jointly eavesdrop on other users, and we propose a linear precoder based on RCI. We use large-system analysis with an approach different from that of \cite{GeraciISWCS11}, and we derive the optimal regularization parameter $\alpha_{\mathrm{LS}}$ and the corresponding achievable secrecy sum-rate. Numerical results confirm the accuracy of the large-system analysis, even when applied to a number of users as low as $4$. Moreover, the RCI precoder with $\alpha_{\mathrm{LS}}$ outperforms several other linear precoding schemes. In fact, it achieves a secrecy sum-rate that has same scaling factor as the sum-rate achieved by the optimum RCI precoder without secrecy requirements. We then propose an iterative power allocation algorithm to obtain the maximum secrecy sum-rate for fixed $\alpha$. We extend our algorithm to maximize the secrecy sum-rate by jointly optimizing the regularization parameter $\alpha$ and the power allocation vector. The proposed power allocation algorithm outperforms RCI with $\alpha_{\mathrm{LS}}$ and equal power allocation (RCI-EP) by up to $20$ percent at practical values of the SNR and for $4$ users and $4$ transmit antennas.

Throughout the paper we use the following notation: bold uppercase (lowercase) letters denote matrices (column vectors); $(\cdot)^{T}$ and $(\cdot)^{\dagger}$ denote matrix transpose and conjugate transpose, respectively; the trace of a matrix is denoted by $\mathrm{tr}\{\cdot\}$, and the Euclidean norm of a vector is indicated by $\| \cdot \|$; $\mathrm{E}[\cdot]$ denotes the expected value of the random variable in the brackets, $\mathcal{CN}(\mu,\sigma^2)$ denotes circularly symmetric complex-Gaussian distribution with mean $\mu$ and variance $\sigma^2$, and we use the notation $[ \cdot ]^+ \stackrel{\triangle}{=} \max(\cdot,0)$.

%% file: Section2.tex
\section{System Model}

We consider the downlink of a narrowband multi-user MIMO system, consisting of a base station (BS) with $M$ antennas which simultaneously transmits $K$ independent confidential messages to $K$ spatially dispersed single-antenna users. Transmission takes place over a block fading channel, where the coherence time of the channel is much longer than one symbol interval. In this model, the transmitted signal is $\mathbf{x} = \left[x_1,\ldots,x_M \right]^{T} \in \mathbb{C}^{M \times 1}$, 
and the received signal at user $k$ is given by
\begin{equation}
y_k=\sum_{j=1}^{M} h_{k,j}x_{j}+n_{k}
\label{eqn:MIMO_scalar}
\end{equation}
where $h_{k,j} \sim \mathcal{CN}(0,1)$ 
is the fading gain between the $j$-th transmit antenna element and the $k$-th user, and $n_{k} \sim \mathcal{CN}(0,\sigma^2)$ is the noise seen at the $k$-th receiver. 
The corresponding vector equation is
\begin{equation}
\mathbf{y}=\mathbf{Hx}+\mathbf{n}
\label{eqn:MIMO_vector}
\end{equation}
where $\mathbf{H} = [h_{k,j}]$ is the $K \times M$ channel matrix, $\mathbf{y} = \left[ y_1, \ldots, y_K \right] ^{T}$ and $\mathbf{n} = \left[ n_1, \ldots, n_K \right] ^{T}$. We impose the long term power constraint E$[ \left\| \mathbf{x} \right\|^{2} ] =1$, assume that E $[ \mathbf{nn^{\dagger}} ] =\sigma^{2} \mathbf{I}$, and define the SNR $\rho = 1/ \sigma ^2$. The transmitted signal $\mathbf{x}$ is obtained at the BS by performing a linear processing on the confidential messages $u_k,~k=1,\ldots,K$. 

It is required that the BS securely transmits each confidential message $u_k$, ensuring that the unintended users receive no information. This is performed at the secrecy rate $R_{s,k}$, defined as follows. Let $\mathrm{Pr}(\mathcal{E}_n)$ be the probability of error at the intended user, $m$ be a confidential message, $\mathbf{y}_e^n$ be the vector of all signals received by the eavesdroppers, and $H(m | \mathbf{y}_e^n)$ be the corresponding equivocation. Then a (weak) secrecy rate $R_{s,k}$ for the intended user is achievable if there exists a sequence of $(2^{nR_{s,k}},n)$ codes such that $\mathrm{Pr}(\mathcal{E}_n) \rightarrow 0$ and $\frac{1}{n}H(m | \mathbf{y}_e^n) \leq \frac{1}{n} H(m) - \varepsilon_n$ with $\varepsilon_n$ approaching zero as $n \rightarrow \infty$ \cite{Khisti10I}.

In general, the behavior of the users cannot be determined by the BS. 
As a worst-case scenario, in our system we assume that for each intended receiver $k$ the remaining $K-1$ users can cooperate to jointly eavesdrop on the message $u_k$.
For each user $k$, the alliance of the $K-1$ cooperating eavesdroppers is equivalent to a single eavesdropper with $K-1$ receive antennas, which is denoted by $\widetilde{k}$.

%% file: Section3.tex
\section{Linear Precoding}

In this section, we derive an achievable secrecy sum-rate for the multi-user MIMO downlink with malicious users by using a linear precoder. Although suboptimal, linear precoding schemes are of particular interest because of their low-complexity implementations and because they can control the amount of crosstalk between the users
\cite{YooJSAC06,Spencer04,Peel05,Joham05}. We then specialize and obtain the secrecy sum-rate achievable by the RCI precoder. RCI is a linear precoding scheme that was proposed to serve multiple users in the multiuser MIMO downlink channel, which has better performance than plain channel inversion, especially at low SNR \cite{Peel05}. 

\subsection{Preliminaries}

In linear precoding, the transmitted vector $\mathbf{x}$ is derived from the vector containing the confidential messages $\mathbf{u} = \left[u_1,\ldots,u_K \right]^{T}$ through a deterministic linear transformation (precoding) \cite{YooJSAC06,Spencer04,Peel05,Joham05}. We assume that the entries of $\mathbf{u}$ are chosen independently, satisfying E$[ \left|u_k\right|^2 ] =1$. We assume spatially homogeneous users, i.e. each user experiences the same received signal power on average, thus the model assumes that their distances from the transmitter are similar.

Let $\mathbf{W} = \left[\mathbf{w}_1,\ldots,\mathbf{w}_K \right]$ be the $M \times K$ precoding matrix, where $\mathbf{w}_k$ is the $k$-th column of $\mathbf{W}$. Then the transmitted signal and the power constraint are, respectively:
\begin{equation}
\mathbf{x} = \frac{1}{\sqrt{\gamma}} \mathbf{Wu} = \frac{1}{\sqrt{\gamma}}\sum_{k=1}^{K} {\mathbf{w}_{k} u_k},
\label{eqn:tx_signal}
\end{equation}
\begin{equation}
\mathrm{E} \left[ \left\| \mathbf{x} \right\| ^2 \right] = \frac{1}{\gamma} \mathrm{E} \left[ \left\| \mathbf{Wu} \right\| ^2 \right] = \frac{1}{\gamma} \sum_{k=1}^{K} \left\| \mathbf{w}_{k} \right\| ^2 = 1,
\label{eqn:power_constraint}
\end{equation}
where $\gamma = \mathrm{tr} \{\mathbf{W ^{\dagger} W} \}$ is the long-term power normalization constant.


\subsection{Achievable Secrecy Sum-Rates with Linear Precoding}

By employing the linear precoding in (\ref{eqn:tx_signal}), the signals observed at receivers $k$ and $\widetilde{k}$ are, respectively
\begin{equation}
\begin{aligned}
y_k &= \frac{1}{\sqrt{\gamma}} \mathbf{h}_{k}^{\dagger} \mathbf{w}_{k} u_k + \frac{1}{\sqrt{\gamma}} \sum_{j \neq k} \mathbf{h}_{k}^{\dagger} \mathbf{w}_{j} u_j + n_{k} \\
\mathbf{y}_{\widetilde{k}} &= \frac{1}{\sqrt{\gamma}} \sum_{k} \mathbf{H}_{\widetilde{k}} \mathbf{w}_{k} u_k + \mathbf{n}_{\widetilde{k}}
\end{aligned}
\label{eqn:wiretap_channel}
\end{equation}
where $\mathbf{n}_{\widetilde{k}} = \left[n_1,\ldots,n_{k-1},n_{k+1},\ldots,n_K \right]^{T}$, $\mathbf{h}_{k}^{\dagger}$ is the $k$-th row of $\mathbf{H}$, and $\mathbf{H}_{\widetilde{k}}$ is a matrix obtained from $\mathbf{H}$ by eliminating the $k$-th row. The channel in (\ref{eqn:wiretap_channel}) is a multi-input, single-output, multi-eavesdropper (MISOME) wiretap channel \cite{Khisti10I}. The transmitter, the intended receiver and the eavesdropper of this MISOME wiretap channel are equipped with $M$, $1$ and $K-1$ virtual antennas, respectively. Due to the simultaneous transmission of the $K$ messages, user $k$ experiences noise and interference from all the $u_j,~j \neq k$.

In the following, we derive an achievable secrecy sum-rate $R_s$ for the multi-user MIMO system with malicious users. Although the design of codes for the multiuser MIMO channel with security constraints is not the focus of this paper, we prove the achievability of $R_s$ with a code construction based on independent codebooks and linear precoding.

\newtheorem{Lemma}{Lemma}
\begin{Lemma}[Codebook construction]
An achievable secrecy sum-rate $R_s$ for the multi-user MIMO system with malicious users is given by
\begin{equation}
R_s \stackrel{\triangle}{=} \sum_{k=1}^{K} {R_{s,k}},
\label{eqn:Rs_linear}
\end{equation}
where $R_{s,k}$ is an achievable secrecy rate for the $k$-th MISOME wiretap channel (\ref{eqn:wiretap_channel}), $k=1,\ldots,K$.
\label{lemma:R_s}
\end{Lemma}
\begin{IEEEproof}
Assume that the BS uses independent codebooks for each user, where each codebook is a code for the scalar wiretap channel \cite{Khisti10I}. The confidential message $u_k$ is obtained as a codeword independently drawn from the code $\mathcal{C}_k$, corresponding to the $k$-th user.
The rate $R_{s,k}$ of the code $\mathcal{C}_k$ is chosen according to the secrecy rate achievable for user $k$ in the presence of eavesdropper $\widetilde{k}$, i.e. by the secrecy rate achievable for the MISOME wiretap channel (\ref{eqn:wiretap_channel}). The existence of such code is guaranteed by the definition of secrecy rate \cite{Csiszar78}.
To construct the vector codeword for the broadcast channel, the scalar codewords for each MISOME wiretap channel are stacked according to $\mathbf{u} = [ u_1,\ldots,u_K]^T$, and no additional binning is required.
The vector $\mathbf{u}$ is then linearly precoded as in (\ref{eqn:tx_signal}), which means that each message $u_k$ is transmitted by beamforming (i.e. signaling with rank one covariance) along the direction $\mathbf{w}_{k}$.
The secrecy sum-rate $R_s$ is then by definition the sum of the simultaneously achievable secrecy rates $R_{s,k}$.
\end{IEEEproof}

\begin{Lemma}
An achievable secrecy rate $R_{s,k}$ for the MISOME wiretap channel (\ref{eqn:wiretap_channel}) is given by
\begin{equation}
R_{s,k} \mathrm{ } = \left[ \log_2 \Big( 1 + \mathrm{SINR}_{k} \Big) - \log_2 \Big( 1 + \mathrm{SINR}_{\widetilde{k}} \Big) \right]^+,
\label{eqn:secrecy_rate}
\end{equation}
where $\mathrm{SINR}_{k}$ and $\mathrm{SINR}_{\widetilde{k}}$ are the signal-to-interference-plus-noise ratios for the message $u_k$ at the intended receiver $k$ and the eavesdropper $\widetilde{k}$, respectively. 
\label{lemma:R_sk}
\end{Lemma}
\begin{IEEEproof}
By noting that the MISOME wiretap channel (\ref{eqn:wiretap_channel}) is a nondegraded broadcast channel \cite{Khisti10I}, the secrecy capacity is given by \cite{Csiszar78}:
\begin{equation}
C_s = \underset{u_k \rightarrow \mathbf{w}_k u_k \rightarrow y_k, \mathbf{y}_{\widetilde{k}}} \max \mathrm{ } {I \Big( u_{k} ; y_{k} \Big) - I \Big( u_{k} ; \mathbf{y}_{\widetilde{k}} \Big)}
\label{eqn:capacity_BCC}
\end{equation}
where $I(x;y)$ denotes mutual information between two random variables $x$ and $y$. The secrecy capacity $C_s$ is given by the difference of the mutual informations at the intended user and at the eavesdropper, respectively. $C_s$ is achieved by maximizing over all joint probability distributions such that a Markov chain $u_k \rightarrow \mathbf{w}_k u_k \rightarrow y_k, \mathbf{y}_{\widetilde{k}}$ is formed, where $u_{k}$ is an auxiliary input variable.
By evaluating (\ref{eqn:capacity_BCC}) with $u_k \sim \mathcal{CN}(0,1)$ and with the linearly precoded data $\mathbf{w}_k u_k$, 
we obtain an achievable secrecy rate $R_{s,k}$ for the MISOME wiretap channel (\ref{eqn:wiretap_channel}) given by 
\begin{equation}
R_{s,k} \mathrm{ } = \left[ I \Big( u_{k} ; y_{k} \Big) - I \Big( u_{k} ; \mathbf{y}_{\widetilde{k}} \Big) \right]^+ \stackrel{a}{=} \left[ I \Big( \mathbf{w}_k u_k; y_{k} \Big) - I \Big( \mathbf{w}_k u_k ; \mathbf{y}_{\widetilde{k}} \Big) \right]^+,
\label{eqn:secrecy_rate_MISOME}
\end{equation}
where (a) follows from $\mathbf{w}_k u_k$ being a deterministic function of $u_{k}$ \cite{Khisti10I}. Equation (\ref{eqn:secrecy_rate}) then follows from (\ref{eqn:secrecy_rate_MISOME}) and from the statistics of $u_k$.
\end{IEEEproof}

From equation (\ref{eqn:secrecy_rate}) it is clearly observed that for high-performance linear precoder design an efficient tradeoff between maximizing $\mathrm{SINR}_{k}$ and minimizing $\mathrm{SINR}_{\widetilde{k}}$ is required.

\newtheorem{Theorem}{Theorem}
\begin{Theorem}
A secrecy sum-rate achievable by multi-user MIMO linear precoding is given by
\begin{equation}
R_{s} = \sum_{k=1}^{K} { \Bigg[ \log_2 \Bigg( 1 + \frac {\left| \mathbf{h}_k^{\dagger} \mathbf{w}_k \right| ^2} {\gamma \sigma^2 + \sum_{j \neq k} {\left| \mathbf{h}_k^{\dagger} \mathbf{w}_j \right| ^2} } \Bigg) - \log_2 \Bigg( 1 + \frac{\left\| \mathbf{H}_{\widetilde{k}} \mathbf{w}_{k} \right\| ^2}{\gamma \sigma^2} \Bigg) \Bigg]^+ }.
\label{eqn:Rs_linear_complete}
\end{equation}
\label{theorem:R_s}
\end{Theorem}
\begin{IEEEproof}
By using Lemma \ref{lemma:R_s} and Lemma \ref{lemma:R_sk}, we have that an achievable secrecy sum-rate is obtained as the sum of the secrecy rates $R_{s,k}$ in (\ref{eqn:secrecy_rate}). A lower bound on the quantities $R_{s,k}$ can be obtained by considering a genie-aided eavesdropper which observes not only the signals $\mathbf{y}_{\widetilde{k}}$ received by its $K-1$ antennas, but also all the confidential messages $u_j, j \neq k$.
Such channel clearly has an achievable secrecy rate smaller than the original channel. 
The genie-aided eavesdropper $\widetilde{k}$ can perform interference cancellation, and it does not see any undesired signal term apart from the received noise $\mathbf{n}_{\widetilde{k}}$.

According to the previous considerations, the signals at the intended receiver and the eavesdropper of the $k$-th equivalent MISOME wiretap channel become, respectively:
\begin{equation}
\begin{aligned}
y_k &= \frac{1}{\sqrt{\gamma}} \mathbf{h}_{k}^{\dagger} \mathbf{w}_{k} u_k + \frac{1}{\sqrt{\gamma}} \sum_{j \neq k} \mathbf{h}_{k}^{\dagger} \mathbf{w}_{j} u_j + n_{k} \\
\mathbf{y}_{\widetilde{k}} &= \frac{1}{\sqrt{\gamma}} \mathbf{H}_{\widetilde{k}} \mathbf{w}_{k} u_k + \mathbf{n}_{\widetilde{k}}
\end{aligned}
\label{eqn:MISOME}
\end{equation}
For the $k$-th equivalent MISOME wiretap channel in (\ref{eqn:MISOME}), the SINRs at the intended user and the eavesdropper are, respectively: 
\begin{equation}
\mathrm{SINR}_{k} = \frac {\left| \mathbf{h}_k^{\dagger} \mathbf{w}_k \right| ^2} {\gamma \sigma^2 + \sum_{j \neq k} {\left| \mathbf{h}_k^{\dagger} \mathbf{w}_j \right| ^2} },
\label{eqn:SINRk_linear}
\end{equation}
\begin{equation}
\mathrm{SINR}_{\widetilde{k}} = \frac{\left\| \mathbf{H}_{\widetilde{k}} \mathbf{w}_{k} \right\| ^2}{\gamma \sigma^2}.
\label{eqn:SINRk_tilde_linear}
\end{equation}
Since the noise in $\mathbf{y}_{\widetilde{k}}$ in (\ref{eqn:MISOME}) is spatially white, the optimal receive filter at $\widetilde{k}$ is the matched filter $( \mathbf{H}_{\widetilde{k}} \mathbf{w}_k ) ^{\dagger}$. Equation (\ref{eqn:SINRk_tilde_linear}) then follows. For a given channel $\mathbf{H}$, substituting (\ref{eqn:SINRk_linear}) and (\ref{eqn:SINRk_tilde_linear}) into (\ref{eqn:secrecy_rate}) and then into (\ref{eqn:Rs_linear}) yields (\ref{eqn:Rs_linear_complete}).
\end{IEEEproof}

For the remainder of the paper, we refer to equation (\ref{eqn:Rs_linear_complete}) as the secrecy sum-rate. We note that it depends on the choice of the precoding matrix $\mathbf{W}$, as well as on the channel $\mathbf{H}$ and the noise variance $\sigma^2$. A possible choice for $\mathbf{W}$, based on regularized channel inversion, is discussed in the following.


\subsection{Achievable Secrecy Sum-Rates with Regularized Channel Inversion}

We now consider RCI precoding for the multi-user MIMO downlink with malicious users. Although CI precoding can achieve secrecy by canceling all signals leaked at the unintended users, this comes at the cost of a poor sum-rate. The RCI precoder has better performance than plain CI, particularly at low SNR \cite{Peel05}. For each message $u_k$, RCI precoding achieves a tradeoff between the signal power at the $k$-th intended user and the crosstalk at the other $(K-1)$ unintended users for each signal. The crosstalk causes interference to the unintended users. In the case when the unintended users are acting maliciously, the crosstalk also causes information leakage. Therefore, RCI achieves a tradeoff between signal power, interference, and information leakage.

With RCI precoding, linear processing exploiting regularization is applied to the vector of messages $\mathbf{u}$ \cite{Peel05}. The RCI precoding matrix is given by
\begin{equation}
\mathbf{W} = \mathbf{H}^{\dagger} \left( \mathbf{H H}^{\dagger} + \alpha \mathbf{I}_K \right) ^{-1}.
\label{eqn:RCI_precoder}
\end{equation}
The transmitted signal $\mathbf{x}$ after RCI precoding can be written as
\begin{equation}
\mathbf{x} = \frac{1}{\sqrt{\gamma}} \mathbf{W} \mathbf{u} = \frac{1}{\sqrt{\gamma}} \mathbf{H}^{\dagger} \left( \mathbf{H H}^{\dagger} + \alpha \mathbf{I}_K \right) ^{-1} \mathbf{u} = \frac{1}{\sqrt{\gamma}} \left( \mathbf{H}^{\dagger} \mathbf{H} + \alpha \mathbf{I}_K \right) ^{-1} \mathbf{H}^{\dagger} \mathbf{u}.
\end{equation}
The latter passes through the channel, producing the vector of received signals
\begin{equation}
\mathbf{y} = \frac{1}{\sqrt{\gamma}} \mathbf{H} \left( \mathbf{H}^{\dagger} \mathbf{H} + \alpha \mathbf{I}_K \right) ^{-1} \mathbf{H}^{\dagger} \mathbf{u} + \mathbf{n}.
\label{eqn:Hs}
\end{equation}
The function of the real nonnegative regularization parameter $\alpha$ is to improve the behavior of the inverse, although it also produces non-zero crosstalk terms in (\ref{eqn:Hs}).

Using RCI precoding, the SINRs (\ref{eqn:SINRk_linear}) and (\ref{eqn:SINRk_tilde_linear}) at the intended user $k$ and the eavesdropper $\widetilde{k}$ become
\begin{equation}
\mathrm{SINR}_{k} = \frac {\left| \mathbf{h}_k^{\dagger} \left( \mathbf{H}^{\dagger} \mathbf{H} + \alpha \mathbf{I}_K \right) ^{-1} \mathbf{h}_k \right| ^2} {\gamma \sigma^2 + \sum_{j \neq k} {\left| \mathbf{h}_k^{\dagger} \left( \mathbf{H}^{\dagger} \mathbf{H} + \alpha \mathbf{I}_K \right) ^{-1} \mathbf{h}_j \right| ^2} },
\label{eqn:SINRk_RCI}
\end{equation}
\begin{equation}
\mathrm{SINR}_{\widetilde{k}} = \frac{\left\| \mathbf{H}_{\widetilde{k}} \left( \mathbf{H}^{\dagger} \mathbf{H} + \alpha \mathbf{I}_K \right) ^{-1} \mathbf{h}_k \right\| ^2}{\gamma \sigma^2},
\label{eqn:SINRk_tilde_RCI}
\end{equation}
where
\begin{equation}
\gamma = \mathrm{tr} \left\{ \mathbf{H}^{\dagger} \mathbf{H} \left( \mathbf{H}^{\dagger} \mathbf{H} + \alpha \mathbf{I}_K \right) ^{-2} \right\}.
\label{eqn:gamma}
\end{equation}
To simplify (\ref{eqn:SINRk_RCI}) and (\ref{eqn:SINRk_tilde_RCI}), we introduce the quantities
\begin{equation}
A_k = \mathbf{h}_k^{\dagger} \left( \mathbf{H}_{\widetilde{k}}^{\dagger} \mathbf{H}_{\widetilde{k}} + \alpha \mathbf{I}_K \right) ^{-1} \mathbf{h}_k \hspace{15pt} \mathrm{and}
\label{eqn:Ak}
\end{equation}
\begin{equation}
B_k = \mathbf{h}_k^{\dagger} \left( \mathbf{H}_{\widetilde{k}}^{\dagger} \mathbf{H}_{\widetilde{k}} + \alpha \mathbf{I}_K \right) ^{-1} \mathbf{H}_{\widetilde{k}}^{\dagger} \mathbf{H}_{\widetilde{k}} \left( \mathbf{H}_{\widetilde{k}}^{\dagger} \mathbf{H}_{\widetilde{k}} + \alpha \mathbf{I}_K \right) ^{-1} \mathbf{h}_k.
\label{eqn:Bk}
\end{equation}
It is then possible to express (\ref{eqn:SINRk_RCI}) as \cite{Nguyen09}
\begin{equation}
\mathrm{SINR}_k = \frac{A_k^2}{B_k + \gamma \sigma^2 \left( 1 + A_k \right)^2}.
\label{eqn:SINRk_Nguyen}
\end{equation}
In a similar fashion, we rewrite (\ref{eqn:SINRk_tilde_RCI}) as
\begin{equation}
\mathrm{SINR}_{\widetilde{k}} = \frac{B_k}{\gamma \sigma^2 \left( 1 + A_k \right)^2}.
\label{eqn:SINRk_tilde_Nguyen}
\end{equation}
By substituting (\ref{eqn:SINRk_Nguyen}) and (\ref{eqn:SINRk_tilde_Nguyen}) into (\ref{eqn:secrecy_rate}) and then into (\ref{eqn:Rs_linear}) we obtain the following expression for the secrecy sum-rates achievable with RCI precoding
\begin{equation}
R_{s} = \sum_{k=1}^{K} { \left[ \log_2 \left( 1 + \frac{A_k^2}{B_k + \gamma \sigma^2 \left( 1 + A_k \right)^2} \right) - \log_2 \left(  1 + \frac{B_k}{\gamma \sigma^2 \left( 1 + A_k \right)^2 }  \right) \right]^+ }.
\label{eqn:Rs_RCI}
\end{equation}

%% file: Section4.tex
\section{Large-System Analysis}

In this section, we consider the performance of the RCI precoder in the large-system regime, where both the number of transmit antennas $M$ and the number of receivers $K$ approach infinity in a fixed ratio. We derive closed-form expressions for the optimal regularization parameter and the optimal secrecy sum-rate achievable with RCI in the large-system regime. We then compare the secrecy sum-rate achieved by the optimized RCI precoder to several other linear precoding schemes. Finally, we evaluate the sum-rate loss due to the secrecy requirements. In this section, we focus on the case $K = M$ because the analysis for this case is more tractable, and it is considered to be most important \cite{Peel05}.

\subsection{Secrecy Sum-Rates in the Large-System Regime}
We define $\xi = \alpha / K$ as the normalized regularization parameter, and note that as $K \rightarrow \infty$, the quantities (\ref{eqn:gamma}), (\ref{eqn:Ak}) and (\ref{eqn:Bk}) converge (almost surely) to \cite{Nguyen09}
\begin{IEEEeqnarray}{rl}
& \lim_{K \rightarrow \infty} \gamma = g \left( \xi \right) + \xi \frac {d} {d \xi} g \left( \xi \right), \\
& \lim_{K \rightarrow \infty} A_k = g \left( \xi \right), \\
& \lim_{K \rightarrow \infty} B_k = g \left( \xi \right) + \xi \frac {d} {d \xi} g \left( \xi \right),
\end{IEEEeqnarray}
respectively, where
\begin{equation}
g \left( \xi \right) = \frac{1}{2} \sqrt{1+ \frac{4}{\xi}} - \frac{1}{2}.
\end{equation}
By substituting the above expressions in (\ref{eqn:SINRk_Nguyen}) and (\ref{eqn:SINRk_tilde_Nguyen}), one can conclude that as $K \rightarrow \infty$, all the SINRs at the intended user $k$ and at the eavesdropper $\widetilde{k}$ converge to a non-random function of the parameter $\xi$ and the noise variance $\sigma^2$. Moreover, these quantities are the same for all confidential messages $u_k$, as $K \rightarrow \infty$. 
Hence, in the large-system regime, it is possible to write the secrecy sum-rate with RCI precoder as
\begin{equation}
R_s \simeq R_{s, \infty} \stackrel{\triangle}{=} K \left[ \log_2 \frac{1+\frac{\rho g(\xi)^2}{\left[ \rho+ \left( 1+g(\xi) \right) ^2 \right] \left[ g(\xi)+ \xi \frac {d} {d \xi} g(\xi) \right] }}{1+\frac{\rho}{ \left( 1+g(\xi) \right) ^2}} \right]^+, \hspace{10pt} \textrm{as } K \rightarrow \infty.
\label{eqn:Rs_large_system}
\end{equation}

\subsection{Selection of the Optimal Regularization Parameter}

The value of the asymptotic secrecy sum-rate $R_{s, \infty}$ in (\ref{eqn:Rs_large_system}) depends on the normalized regularization parameter $\xi$. We now derive the optimal value $\xi_{\mathrm{opt}}$ that maximizes $R_{s, \infty}$. 

\begin{Lemma}
The optimal normalized regularization parameter in the large-system regime is given by
\begin{equation}
\xi_{\mathrm{opt}}
= \frac{1}{3 \rho + 1 + \sqrt{3 \rho + 1}}.
\label{eqn:xi_opt}
\end{equation} 
\label{lemma:xi_opt}
\end{Lemma}
\begin{IEEEproof}
The value of $\xi_{\mathrm{opt}}$ is obtained as the stationary point of the secrecy sum-rate $R_{s, \infty}$, which can be found by setting to zero the derivative of the logarithm in (\ref{eqn:Rs_large_system}), by applying some algebraic manipulations, and showing that the maximum value is nonnegative.
\end{IEEEproof}

As in the case with no secrecy requirements, the value of $\xi_{\mathrm{opt}}$ is a function of the SNR $\rho$. In a multi-user channel without secrecy requirements, the choice $\xi = 1 / \rho$ is optimal for large $K$, as it maximizes the sum-rate of the system \cite{Peel05}. However, this value is no longer optimal in a multi-user channel with malicious users. In fact, because of the secrecy requirements the crosstalk terms appear twice in the secrecy sum-rate expression (\ref{eqn:Rs_linear_complete}). As a consequence, $\xi = 1 /\rho$ is too large and gives too much crosstalk to the other users. This was proven in \cite{GeraciISWCS11} and it is also easily confirmed by the following inequality:
\vspace{-0.2cm}
\begin{equation}
\xi_{\mathrm{opt}} < \frac{1}{3 \rho} < \frac{1}{\rho} \hspace{10pt} \forall \rho. 
\end{equation}

Similarly to the case with no secrecy requirements, $\xi_{\mathrm{opt}}$ decreases as we increase $\rho$. The high-SNR asymptote of $\xi_{\mathrm{opt}}$ is given by
\vspace{-0.2cm}
\begin{equation}
\xi_{\mathrm{opt}} \simeq \frac {1} {3 \rho}, \hspace{10pt} \textrm{as } \rho \rightarrow \infty 
\label{eqn:alpha_opt_large_SNR}
\end{equation}
and $\xi_{\mathrm{opt}}$ tends to zero if $\rho \rightarrow \infty$.

Unlike the case with no confidentiality, the optimum normalized regularization parameter is upper bounded and it does not tend to infinity as $\rho$ tends to zero. The low-SNR asymptote of $\xi_{\mathrm{opt}}$ is 
\begin{equation}
\xi_{\mathrm{opt}} = \frac {1} {2},  \hspace{10pt} \textrm{for } \rho = 0.
\end{equation}

In the remainder of the paper, we will denote by $\alpha_{\mathrm{LS}} = K \xi_{\mathrm{opt}}$ the unnormalized large-system regularization parameter.

\subsection{Optimal Secrecy Sum-Rate}

It is now possible to obtain an expression for the optimal secrecy sum-rate of the RCI precoder in the large-system regime. The optimal secrecy sum-rate is a function of the SNR $\rho$ and the number of users $K$ only. 

\begin{Theorem}
The optimal secrecy sum-rate $R_{s,\infty}^{\mathrm{RCI}}$ achievable by the RCI precoder in the large-system regime is given by
\vspace{-0.2cm}
\begin{equation}
R_{s,\infty}^{\mathrm{RCI}} \stackrel{\triangle}{=} \max_{\xi} R_{s, \infty} = K \log_2 \frac {9 \rho + 2 + \left(6 \rho + 2 \right) \sqrt{3 \rho + 1}} {4 \left( 4 \rho + 1 \right)}.
\label{eqn:Rs_opt}
\end{equation}
\label{theorem:Rs_opt}
\end{Theorem}
\begin{IEEEproof}
Equation (\ref{eqn:Rs_opt}) is obtained by substituting (\ref{eqn:xi_opt}) in (\ref{eqn:Rs_large_system}) and applying some algebraic manipulations.
\end{IEEEproof}

The secrecy sum-rate $R_{s,\infty}^{\mathrm{RCI}}$ in (\ref{eqn:Rs_opt}) satisfies
\begin{equation}
R_{s,\infty}^{\mathrm{RCI}} > 0 \quad \forall \rho > 0,
\end{equation}
and the high-SNR asymptote of $R_{s,\infty}^{\mathrm{RCI}}$ is given by
\begin{equation}
R_{s,\infty}^{\mathrm{RCI}} \simeq \frac{K}{2} \log_2 {\frac{27}{64}} + \frac{K}{2} \log_2 {\rho}, \hspace{10pt} \textrm{as } \rho \rightarrow \infty.
\label{eqn:Rs_opt_high_SNR}
\end{equation}
Therefore, in the large-system regime the secrecy sum-rate for optimal $\xi$ scales logarithmically with high SNR, and it scales linearly as $K/2$ with the number of users.

We have shown that although for $K \rightarrow \infty$ the number of eavesdroppers $K-1$ for each message tends to infinity, a positive secrecy sum-rate is still achievable. This occurs because the number of transmit antennas $M = K$ also tends to infinity, and it is larger than the number of eavesdroppers. Therefore, for each message the transmitter is able to control the amount of interference and information leakage.

We now compare the per-user secrecy rate achieved by RCI to the secrecy capacity of the MISOME channel, $C_{s}^{\mathrm{MISOME}}$, in the high-SNR regime. The former is obtained by dividing (\ref{eqn:Rs_opt_high_SNR}) by the number of users $K$, and it can be further approximated by
\begin{equation}
\frac{R_{s,\infty}^{\mathrm{RCI}}}{K} \simeq \frac{1}{2} \log_2 {\rho}, \hspace{10pt} \textrm{as } \rho \rightarrow \infty.
\label{eqn:Rs_per_user}
\end{equation}
The value of $C_{s}^{\mathrm{MISOME}}$ was obtained in \cite{Khisti10I}, and it can be approximated by the following lower bound
\begin{equation}
C_{s}^{\mathrm{MISOME}} \geq \frac{1}{2} \log_2 {\rho}, \hspace{10pt} \textrm{as } \rho \rightarrow \infty,
\label{eqn:Cs_Khisti}
\end{equation}
which is tight at high SNR \cite{Khisti10I}.
We note that in $C_{s}^{\mathrm{MISOME}}$ from \cite{Khisti10I} a single-user system is considered. Therefore, only one message is transmitted to one legitimate user, and the user does not experience any interference. For large SNR, the RCI precoder achieves a per-user secrecy rate which is the same as the secrecy capacity of a single-user system.

\subsection{Comparison to Other Linear Schemes}

In the following, we compare the secrecy sum-rate in ($\ref{eqn:Rs_opt}$) achieved by the RCI precoder with $\xi_{\mathrm{opt}}$ to the secrecy sum-rates obtained from ($\ref{eqn:Rs_large_system}$) by using: 1) $\xi = 0$ (CI precoder), 2) $\xi \rightarrow \infty$ (matched-filter precoder) and 3) $\xi = 1 / \rho$ (optimum RCI precoder without secrecy requirements).

The aim of the CI precoder is to cancel all the interference and information leakage, therefore yielding to a secrecy sum-rate that coincides with the sum-rate. We note that for the CI precoder it is $\xi=0$, and the precoding matrix is given by
\begin{equation}
\mathbf{W} = \mathbf{H}^{\dagger} \left( \mathbf{H H}^{\dagger} \right) ^{-1}.
\label{eqn:W_ZF}
\end{equation}
In order for the inverse in (\ref{eqn:W_ZF}) to exist, it is required that $K \leq M$. 

The secrecy sum-rate achieved by CI in the large-system regime grows at most sublinearly with $K \rightarrow \infty$. In fact,
\begin{equation}
\lim_{\xi \rightarrow 0} \lim_{K \rightarrow \infty} \frac{R_s}{K} = 0.
\label{eqn:Rs_ZF}
\end{equation}
This result is consistent with \cite{Nguyen09}, where it was shown that the CI precoder performs poorly in the large-system regime when the number of antennas equals the number of users.

Similarly, we calculate the secrecy sum-rate achieved when $\xi \rightarrow \infty$ (matched-filter precoding). Here, the transmitter beamforms in a direction such as to maximize the signal strenght of each user, without taking into account the interference it creates and the amount of resulting information leakage.
The secrecy sum-rate achieved by matched-filter precoding in the large-system regime is zero. In fact,
\begin{equation}
\lim_{\xi \rightarrow \infty} \lim_{K \rightarrow \infty} \frac{R_s}{K} = \left[ \log_2 \frac{2 \rho + 1}{ \left( \rho + 1 \right)^2} \right]^+ = 0.
\label{eqn:Rs_MF}
\end{equation}
Clearly, matched-filter precoding performs poorly compared to the optimal RCI precoder. This is due to the intended user suffering from a large amount of interference, while the eavesdroppers may cancel the interference by cooperating.

Finally, we consider $\xi = 1 / \rho$, which is the value that maximizes the sum-rate of the system without secrecy requirements \cite{Peel05}.
The secrecy sum-rate $R_{s,\infty}^{\circ}$ achieved by RCI with $\xi = 1 / \rho$ in the large-system regime is given by
\begin{equation}
R_{s,\infty}^{\circ} = K \log_2 \frac{4 \rho + 1 + \left( 2 \rho + 1 \right) \sqrt{4 \rho + 1}}{2 \left( 4 \rho + 1 \right)}.
\end{equation}
We observe that the RCI scheme with $\xi = 1/ \rho$ outperforms the CI and the matched-filter precoding schemes in the large-system regime, but it is suboptimal compared to the use of $\xi_{\mathrm{opt}}$.
For high SNR, the per-antenna secrecy sum-rate gain provided by using $\xi = \xi_{\mathrm{opt}}$ in place of $\xi = 1 / \rho$ is given by
\begin{equation}
\lim_{\rho \rightarrow \infty} \frac{ R_{s,\infty}^{\mathrm{RCI}} - R_{s,\infty}^{\circ}}{K} = \log_2 \frac{3 \sqrt{3}}{4} \approx 0.38 \hspace{5pt} \mathrm{bits}.
\label{eqn:Peel_gap_large_SNR}
\end{equation}

\subsection{Secrecy Loss}

We now consider the \emph{secrecy loss}, i.e. the sum-rate loss due to the secrecy requirements. We define this as the difference between the optimal sum-rate $R_{\infty}^{\circ}$ without secrecy requirements and the secrecy sum-rate $R_{s,\infty}^{\mathrm{RCI}}$ in ($\ref{eqn:Rs_opt}$). The sum-rate $R_{\infty}^{\circ}$ is obtained with RCI and $\xi = 1/ \rho$, and it is given by \cite{Nguyen09}
\begin{equation}
R_{\infty}^{\circ} = K \log_2 \frac{1 + \sqrt{4 \rho + 1}}{2}.
\label{eqn:R_Peel}
\end{equation}
The high-SNR asymptote of the sum-rate in (\ref{eqn:R_Peel}) is
\begin{equation}
R_{\infty}^{\circ} \simeq \frac{K}{2} \log_2 \rho, \hspace{10pt} \textrm{as } \rho \rightarrow \infty.
\label{eqn:R_Peel_large_SNR}
\end{equation}
For high SNR, the per-antenna secrecy loss is given by
\begin{equation}
\lim_{\rho \rightarrow \infty} \frac{R_{\infty}^{\circ} - R_{s,\infty}^{\mathrm{RCI}}}{K} = \frac{1}{2} \log_2 \frac{64}{27} \approx 0.62 \hspace{5pt} \mathrm{bits}.
\label{eqn:secrecy_loss_large_SNR}
\end{equation}
By comparing (\ref{eqn:R_Peel_large_SNR}) to (\ref{eqn:Rs_opt_high_SNR}), one can conclude that the secrecy requirements do not change the linear scaling factor for large SNR. In other words, the RCI precoder with $\xi_{\mathrm{opt}}$ achieves a secrecy sum-rate that has same scaling factor $K/2$ as the sum-rate achieved by the optimum RCI precoder without secrecy requirements in \cite{Peel05}. The RCI precoder with $\xi_{\mathrm{opt}}$ can achieve secrecy with a penalty in terms of the per-antenna sum-rate given by (\ref{eqn:secrecy_loss_large_SNR}). The secrecy loss (\ref{eqn:secrecy_loss_large_SNR}) corresponds to a power loss of a factor $64/27 \approx 3.75$dB. Therefore, the RCI precoder with $\xi_{\mathrm{opt}}$ can achieve secrecy without reducing the sum-rate of the system, as long as the transmitted power is increased by $3.75$dB.

%% file: Section5.tex
\section{Power Allocation}

In this section, we consider power allocation for the RCI precoder. We first propose a new algorithm to obtain the power allocation vector $\mathbf{p}$ which achieves the optimal secrecy sum-rate with a fixed regularization parameter $\alpha$. We then extend our algorithm to jointly optimize $\mathbf{p}$ 
and $\alpha$.

\subsection{Achievable Secrecy Sum-Rates}
We consider the RCI precoding matrix with arbitrary power allocation given by
\begin{align}
\mathbf{W}_{\mathrm{p}} = \mathbf{WD} = \mathbf{H}^\dag(\mathbf{HH}^\dag + \alpha\mathbf{I})^{-1}\mathbf{D},
\label{eqn:PA_precoder}
\end{align}
where $\mathbf{D} = \mathrm{diag}(\sqrt{\mathbf{p}})$, and $\mathbf{p} = [p_1,\ldots,p_K]^T$ is the power allocation vector. The vector $\mathbf{p}$ must be chosen such that the power constraint $\mathrm{tr}\left\{\mathbf{W}_{\mathrm{p}}^\dag \mathbf{W}_{\mathrm{p}} \right\} = 1$ is met. Clearly, (\ref{eqn:PA_precoder}) generalizes the RCI precoder $\mathbf{W}$ with equal power allocation (RCI-EP) in (\ref{eqn:RCI_precoder}).

When the precoder $\mathbf{W}_{\mathrm{p}}$ is used, the SINR at the $k$-th intended user, given by (\ref{eqn:SINRk_linear}),  becomes
\begin{equation}
\mathrm{SINR}_k = \frac{p_k|\textbf{h}_k^{\dagger} \textbf{w}_k|^2}{\sum_{j\neq k}p_j|\textbf{h}_k^{\dagger} \textbf{w}_j|^2 + \sigma^2},
\label{eqn:SINRk_PA}
\end{equation}
and the SINR at the eavesdropper $\widetilde{k}$, given by (\ref{eqn:SINRk_tilde_linear}), becomes
\begin{equation}
\mathrm{SINR}_{\widetilde{k}} = \frac{p_k \|\mathbf{H}_{\widetilde{k}} \mathbf{w}_k\|^2}{\sigma^2} = \frac{p_k\sum_{j\neq k} |\textbf{h}_j^{\dagger} \textbf{w}_k|^2}{\sigma^2}.
\label{eqn:SINRk_tilde_PA}
\end{equation}
From (\ref{eqn:SINRk_PA}) and (\ref{eqn:SINRk_tilde_PA}), we obtain the achievable secrecy sum-rate with power allocation
\begin{equation}
R_{s}^{\mathrm{PA}} = \sum_{k=1}^{K} { \Bigg[ \log_2 \Bigg( 1 + \frac{p_k|\textbf{h}_k^{\dagger} \textbf{w}_k|^2}{\sum_{j\neq k}p_j|\textbf{h}_k^{\dagger} \textbf{w}_j|^2 + \sigma^2} \Bigg) - \log_2 \Bigg( 1 + \frac{p_k\sum_{j\neq k} |\textbf{h}_j^{\dagger} \textbf{w}_k|^2}{\sigma^2} \Bigg) \Bigg]^+ }.
\label{eqn:secrecy_sum_rate_PA}
\end{equation}

\subsection{Power Control}

To obtain the optimal power allocation vector $\mathbf{p}$, we are required to solve the non-convex optimization problem
\begin{align}
\begin{aligned}
& \underset{\textbf{p}}{\text{maximize}}
& & R_s^{\mathrm{PA}}(\mathbf{p}) \\
& \text{subject to}
& & \mathrm{tr}\left\{\mathbf{W}_{\mathrm{p}}^\dag \mathbf{W}_{\mathrm{p}}\right\} \leq 1,
\end{aligned}
\label{poweropt0}
\end{align}
where $R_s^{\mathrm{PA}}(\mathbf{p})$ is given by (\ref{eqn:secrecy_sum_rate_PA}), $\mathbf{W}_{\mathrm{p}}$ is given by (\ref{eqn:PA_precoder}), and the maximum total transmit power over all antennas is one. In the following, we will ignore the notation $[ \cdot ]^+$ in (\ref{eqn:secrecy_sum_rate_PA}) in the maximization problem. In fact, any negative term in the sum can be replaced by zero (thus increasing the sum) by using $p_k = 0$ which is always feasible.

We now reformulate the problem (\ref{poweropt0})
by applying the transformation $\widetilde{p}_k = \log p_k,~k=1\ldots,K$, and obtain the optimization problem
\begin{align}
\begin{aligned}
& \underset{\mathbf{\widetilde{p}}} {\text{maximize}} 
& & R_s^{\mathrm{PA}}(\widetilde{\mathbf{p}}) \\
& \text{subject to}
& & \mathrm{tr}\left\{\mathbf{W}_{\mathrm{p}}^\dag \mathbf{W}_{\mathrm{p}} \right\} \leq 1,
\end{aligned}
\label{poweropt}
\end{align}
where $\widetilde{\mathbf{p}} = [\widetilde{p}_1,\ldots,\widetilde{p}_K]^T$. 

\begin{Lemma}
The second term of the objective function, $R_s^{\mathrm{PA}}(\widetilde{\mathbf{p}})$, of (\ref{poweropt}) is concave.
\label{T2}
\end{Lemma}
\begin{IEEEproof}
The second term and its first and second derivatives are
\begin{align}
\begin{aligned}
- \log_2 \left( 1 \! + \! \mathrm{SINR}_{\widetilde{k}} \right) & = - \log_2\left(1 \! + \! \frac{e^{\widetilde{p}_k} \sum_{j\neq k} |\textbf{h}_j^{\dagger}\textbf{w}_k|^2}{\sigma^2}\right), \\
- \frac{\partial \log_2 \left( 1 + \mathrm{SINR}_{\widetilde{k}} \right)}{\partial\widetilde{p}_k} & = - \frac{ e^{\widetilde{p}_k}\sum_{j\neq k} |\textbf{h}_j^{\dagger}\textbf{w}_k|^2}{\sigma^2 + e^{\widetilde{p}_k}\sum_{j\neq k} |\textbf{h}_j^{\dagger}\textbf{w}_k|^2} \log_2 e, \\
- \frac{\partial^2 \log_2 \left( 1 + \mathrm{SINR}_{\widetilde{k}} \right)}{\partial\widetilde{p}_k^2} & = - \frac{e^{\widetilde{p}_k}\sum_{j\neq k} |\textbf{h}_j^{\dagger}\textbf{w}_k|^2 \sigma^2}{\left(\sigma^2 + e^{\widetilde{p}_k}\sum_{j\neq k} |\textbf{h}_j^{\dagger}\textbf{w}_k|^2\right)^2} \log_2 e \leq 0.
\end{aligned}
\end{align}
Hence, by the second order condition [30, \textsection 3.4.3]
, $-\log_2 \left( 1 + \mathrm{SINR}_{\widetilde{k}} \right)$ is concave.
\end{IEEEproof}

In order to solve the problem (\ref{poweropt}), we consider a modified version of the method as in \cite{Papandriopoulos08} and \cite{SungICC10} which is based on a reformulation of (\ref{poweropt}). This approach guarantees an improvement in the performance over the standard high-SNR approximation in fading channels \cite{SungICC10}. In order to obtain the reformulation, we use the following bound obtained in \cite{Papandriopoulos08}
\begin{equation}
\label{lowbound}
a\log z  + b \leq \log(1 + z), \hspace{10pt} a = \frac{z_0}{1 + z_0} \hspace{10pt} \textrm{and} \hspace{10pt} b = \log(1 + z_0) - \frac{z_0}{1 + z_0}\log z_0,
\end{equation}
for some $z_0 \geq 0$, with equality when $z = z_0$.

\begin{Lemma}
With the change of variables $\widetilde{p}_k = \log p_k,~k=1\ldots,K$, the lower bound
\begin{align}
\frac{a_k}{\log 2} \log \left( \frac{e^{\widetilde{p}_k}|\textbf{h}_k^{\dagger}\textbf{w}_k|^2}{\sum_{j\neq k}e^{\widetilde{p}_j}|\textbf{h}_k^{\dagger}\textbf{w}_j|^2 + \sigma^2} \right) + \frac{b_k}{\log 2} \leq \log_2 \left( 1 + \frac{p_k|\textbf{h}_k^{\dagger} \textbf{w}_k|^2}{\sum_{j\neq k}p_j|\textbf{h}_k^{\dagger} \textbf{w}_j|^2 + \sigma^2} \right),
\label{eqn:lower_bound}
\end{align}
is concave in $\widetilde{p}_k,~k=1,\ldots,K$.
\label{T3}
\end{Lemma}
\begin{IEEEproof}
The result follows immediately using the method in Lemma~\ref{T2}.
\end{IEEEproof}

We showed in Lemma~\ref{T2} that the second term of (\ref{eqn:secrecy_sum_rate_PA}) is concave by the second order condition. By using the lower bound in (\ref{eqn:lower_bound}) for the first term of (\ref{eqn:secrecy_sum_rate_PA}), we obtain a concave objective function. Since the constraints are affine, the optimization problem arising from (\ref{poweropt}) and the bound (\ref{eqn:lower_bound}) is a convex optimization problem. This convex optimization problem is given by
\begin{align} \label{poweroptconv}
\begin{aligned}
& \underset{\mathbf{\widetilde{p}}}{\text{maximize}} & & \sum_{k=1}^K\left[\frac{a_k}{\log 2} \log \! \left( \frac{e^{\widetilde{p}_k}|\textbf{h}_k^{\dagger}\textbf{w}_k|^2}{\sum_{j\neq k} e^{\widetilde{p}_j} |\textbf{h}_k^{\dagger}\textbf{w}_j|^2 \! + \! \sigma^2} \right) \! + \! \frac{b_k}{\log 2}
\! - \! \log_2 \! \left(1 \! + \! \frac{e^{\widetilde{p}_k}\sum_{j\neq k} |\textbf{h}_j^{\dagger}\textbf{w}_k|^2}{\sigma^2}\right)\right] \\
& \text{subject to} & & \mathrm{tr}\left\{\mathbf{W}_{\mathrm{p}}^\dag \mathbf{W}_{\mathrm{p}}\right\} \leq 1
\end{aligned}
\end{align}

The power allocation vector can then be obtained using Algorithm 1 in Table \ref{algorithms}. To show that Algorithm 1 converges monotonically to a local optimum, we note the constraint is the same for both the $t$-th and $(t+1)$-th subproblems. Hence, the solution of the $t$-th subproblem (\ref{poweroptconv}) is also feasible for the $(t+1)$-th subproblem (\ref{poweroptconv}). Moreover, by the bound in (\ref{lowbound}), the objective function is monotonically increasing and converges to a local optimum.

\subsection{Proposed Precoding Scheme}

Having established an algorithm to determine the optimal power allocation vector $\mathbf{p}$ for a fixed $\alpha$, we now obtain our precoding scheme by considering the joint optimization of $\alpha$ and $\mathbf{p}$. The joint optimization problem can be written as
\begin{align}\label{poweropt1}
\begin{aligned}
& \underset{\textbf{p},\alpha}{\text{maximize}}
& & R_s^{\mathrm{PA}}(\mathbf{p},\alpha) \\
& \text{subject to}
& & \mathrm{tr}\left\{ \mathbf{W}_{\mathrm{p}}^\dag \mathbf{W}_{\mathrm{p}} \right\} \leq 1.
\end{aligned}
\end{align}
Even after using the transformation $\widetilde{p}_k = \log p_k,~k=1,\ldots,K$, the problem (\ref{poweropt1}) is non-convex. To solve this problem, we propose Algorithm 2 in Table \ref{algorithms}. 

At each iteration, Algorithm 2 optimizes the regularization parameter $\alpha$ and subsequently the power allocation vector $\mathbf{p}$. It is straightforward to prove that Algorithm 2 converges monotonically and it thus provides with a locally optimal pair $(\alpha,\mathbf{p})$ for the proposed linear precoder. In Section VI we show via simulations that the proposed precoder with jointly optimal regularization parameter and power allocation vector outperforms RCI precoding with $\alpha_{\mathrm{LS}}$ and equal power allocation (RCI-EP). 

\begin{table}[!t]
\renewcommand{\arraystretch}{1.3}
\caption{Proposed algorithms for power allocation.}
\label{algorithms}
\centering
\begin{tabular}{|p{6.5cm}|p{9.1cm}|}
\hline
\centering
\vspace{3pt}
\textbf{Algorithm 1}
\vspace{3pt}
&
\centering
\vspace{3pt}
\textbf{Algorithm 2}
\vspace{3pt}
\tabularnewline
\hline
\renewcommand{\labelitemi}{$$}
\begin{itemize}
\item Initialize iteration counter $t = 0$
\item Initialize all $a_k^{(t)} = 1$, $b_k^{(t)} = 0$
\item \textbf{repeat}
\item ~~~Solve (\ref{poweroptconv}) to obtain $\mathbf{\widetilde{p}}^{(t)}$
\item ~~~Update $a_k^{(t)},~b_k^{(t)}$ at $z_0 = \mathrm{SINR}_k(\mathbf{\widetilde{p}}^{(t)})$
\item ~~~Increment $t$
\item \textbf{until} convergence
\item Obtain $p_k = e^{\widetilde{p}_k},~k = 1, \ldots, K$
\end{itemize}
&
\renewcommand{\labelitemi}{$$}
\begin{itemize}
\item Initialize iteration counter $t_1 = 0,~t_2 = 0$
\item Initialize $p_k = 1/\gamma$, and set $\widetilde{p}_k = \log p_k,~k = 1, \ldots, K$
\item Initialize $\alpha_0 = K \xi_{\mathrm{opt}}$ using equation (\ref{eqn:xi_opt})
\item \textbf{repeat}
\item ~~~Increment $t_1$
\item ~~~Obtain $\alpha^*_{t_1}$ using steepest descent with $\alpha_{t_1-1}$ as initial point
\item ~~~Initialize all $a_k^{(t_2)} = 1$, $b_k^{(t_2)} = 0$
\item ~~~\textbf{repeat}
\item ~~~~~~Solve (\ref{poweroptconv}) to obtain $\mathbf{\widetilde{p}}^{(t_2)}$
\item ~~~~~~Update $a_k^{(t_2)},~b_k^{(t_2)}$ at $z_0 = \mathrm{SINR}_k(\alpha^*_{t_1},\mathbf{\widetilde{p}}^{(t_2)})$
\item ~~~~~~Increment $t_2$
\item ~~~\textbf{until} convergence
\item ~~~Set $\mathbf{\widetilde{p}}=\mathbf{\widetilde{p}}^{(t_2)}$
\item \textbf{until} convergence
\item Obtain $p_k = e^{\widetilde{p}_k},~k = 1, \ldots, K$
\label{locit}
\end{itemize}
\tabularnewline
\hline
\end{tabular}
\end{table}

%% file: Section6.tex
\section{Numerical Results}

\begin{figure}
\centering
\includegraphics[width=\columnwidth]{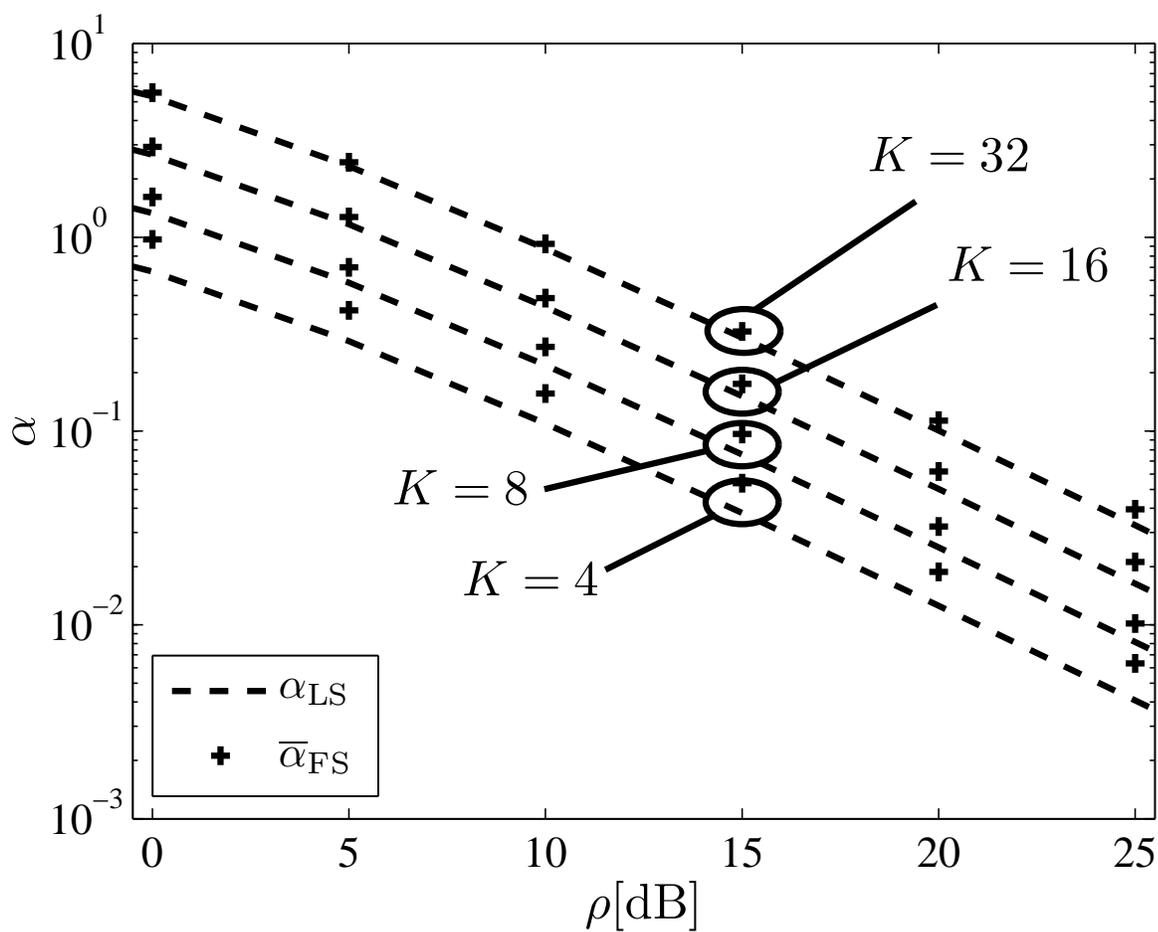}
\caption{Comparison between the large-system regularization parameter $\alpha_{\mathrm{LS}}$ and the value $\overline{\alpha}_{\mathrm{FS}}$ that maximizes the average secrecy sum-rate for finite $K$.}
\label{fig:alpha_formula_vs_sim}
\end{figure}

\begin{figure}
\centering
\includegraphics[width=\columnwidth]{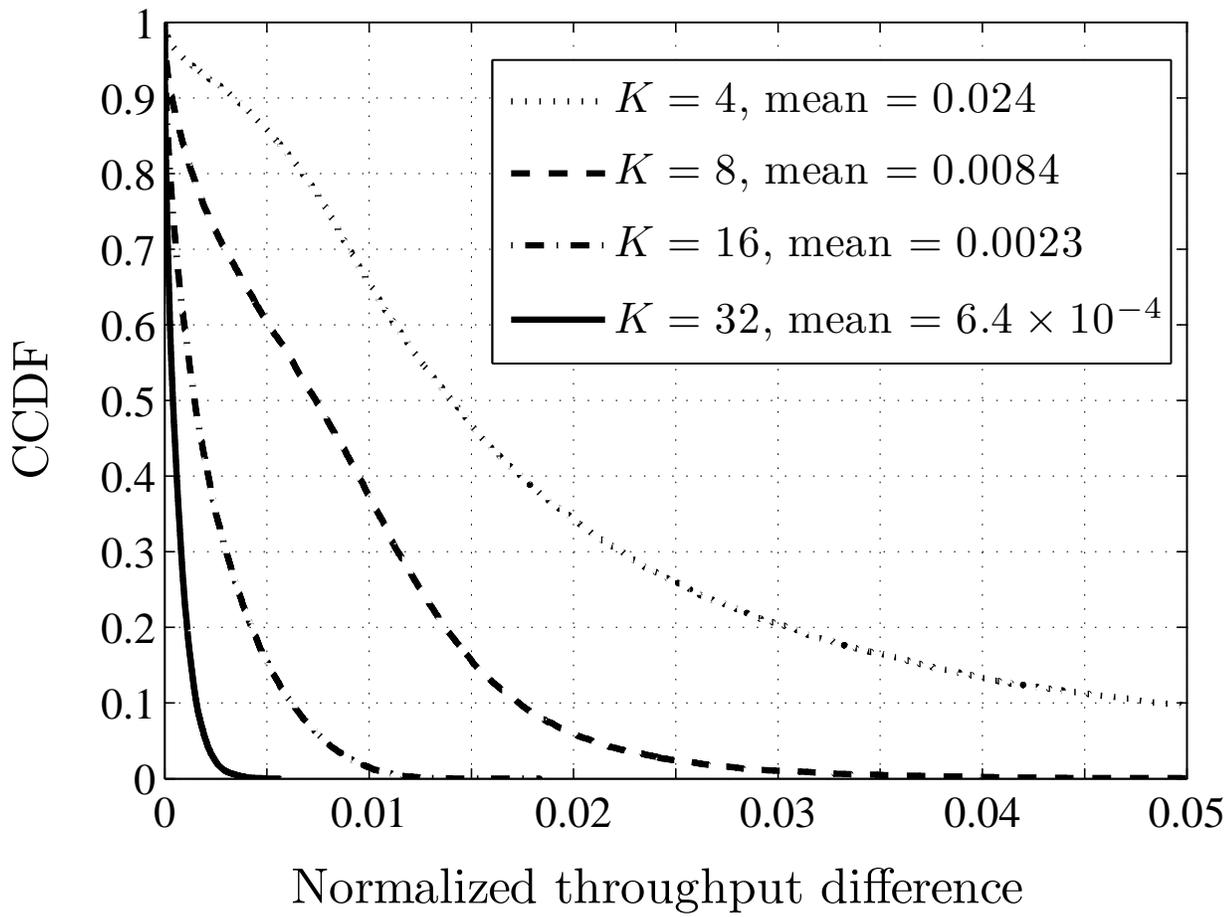}
\caption{Complementary cumulative distribution function (CCDF) of the normalized secrecy sum-rate difference between using $\alpha_{\mathrm{FS}}(\mathbf{H})$ and $\alpha_{\mathrm{LS}}$ with $\rho = 10$dB.}
\label{fig:throughput_difference}
\end{figure}

\begin{figure}
\centering
\includegraphics[width=\columnwidth]{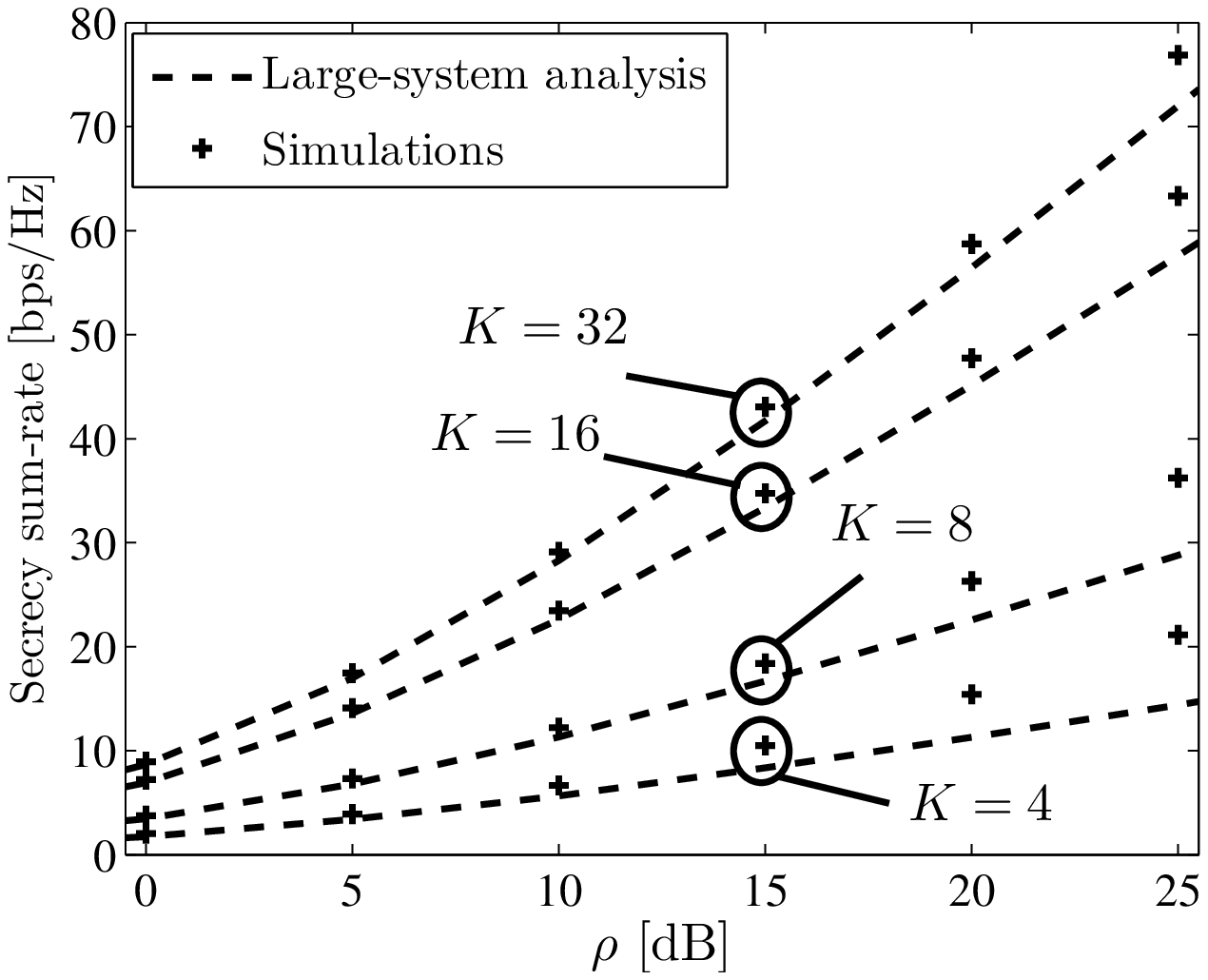}
\caption{Comparison between the secrecy sum-rate in the large-system regime (\ref{eqn:Rs_opt}) and the simulated secrecy sum-rate for finite $K$.}
\label{fig:secrecy_sum_rate_formula_vs_sim}
\end{figure}

In this section, we show the performance of our proposed precoding scheme via simulations. We also consider the finite user scenario to show that many results from the large-system analysis in Section IV hold for a small number of users. The precoding matrix $\mathbf{W}$ was normalized by $\sqrt{\gamma}$, as in (\ref{eqn:gamma}), in order to meet the power constraint in (\ref{eqn:power_constraint}). This corresponds to a long-term power constraint, which does not require the receivers to know the instantaneous value of $\gamma$ \cite{Peel05}. In the following, we denote by $\alpha_{\mathrm{LS}} = K \xi_{\mathrm{opt}}$ the large-system regularization parameter, obtained from (\ref{eqn:xi_opt}).

Fig. \ref{fig:alpha_formula_vs_sim} compares the large-system regularization parameter $\alpha_{\mathrm{LS}}$ to the optimal regularization parameter $\overline{\alpha}_{\mathrm{FS}}$ for a finite number of users. The value of $\overline{\alpha}_{\mathrm{FS}}$ was found by using single-variable numerical optimization to maximize the mean value of the secrecy sum-rate in (\ref{eqn:Rs_RCI}). The figure shows the finite-system and large-system regularization parameters at practical SNR values for four different numbers of users: $4$, $8$, $16$ and $32$. We observe that as the number of users $K$ increases, the value of $\overline{\alpha}_{\mathrm{FS}}$ approaches the large-system regularization parameter $\alpha_{\mathrm{LS}}$.

In Fig. \ref{fig:throughput_difference} we demonstrate that using the large-system regularization parameter $\alpha_{\mathrm{LS}}$ in a finite-size system does not cause a significant loss in the secrecy sum-rate compared to using a regularization parameter $\alpha_{\mathrm{FS}}(\mathbf{H})$ optimized for each channel realization. The figure shows the complementary cumulative distribution function (CCDF) of the normalized secrecy sum-rate difference between using $\alpha_{\mathrm{LS}}$ and $\alpha_{\mathrm{FS}}(\mathbf{H})$ as the regularization parameter of the RCI precoder for $K = 4$, $8$, $16$, $32$ users at an SNR of $10$dB. The difference is normalized by dividing by the secrecy sum-rate of the precoder that uses $\alpha_{\mathrm{FS}}(\mathbf{H})$. We observe that the average normalized secrecy sum-rate difference is less than $2.4$ percent for all values of $K$. As a result, the large-system regularization parameter $\alpha_{\mathrm{LS}}$ may be used instead of the finite-system regularization parameter with only a small loss of performance. Moreover, the value of $\alpha_{\mathrm{LS}}$ does not need to be calculated for each channel realization.

Fig. \ref{fig:secrecy_sum_rate_formula_vs_sim} compares the analytical secrecy sum-rate of the RCI precoder in (\ref{eqn:Rs_opt}) to the simulated secrecy sum-rate of the RCI precoder with a finite number of users, which is averaged over $10^3$ channels. The RCI precoder with a finite number of users was obtained by using the regularization parameter $\overline{\alpha}_{\mathrm{FS}}$, found by simulation, that maximizes the average secrecy sum-rate. We observe that the large-system analysis is accurate at low SNR for all values of $K$. Moreover as $K$ increases, the large-system analysis is accurate for larger values of the SNR.  

\begin{figure}[t]
\centering
\subfigure[$K=4$]{
   \includegraphics[width=3.1in]{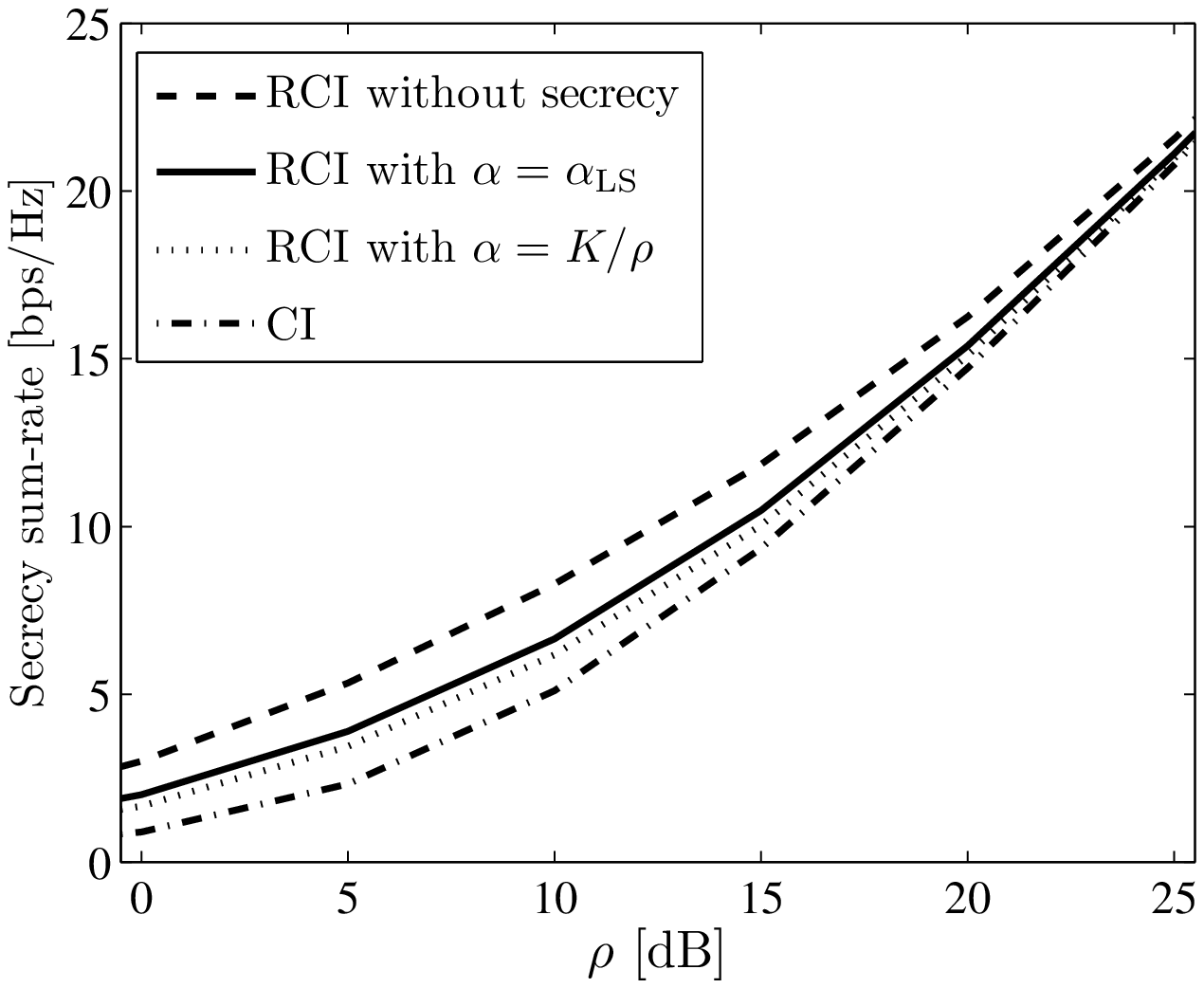}
   \label{fig:all_schemes_4}
 }
\subfigure[$K=8$]{
   \includegraphics[width=3.1in]{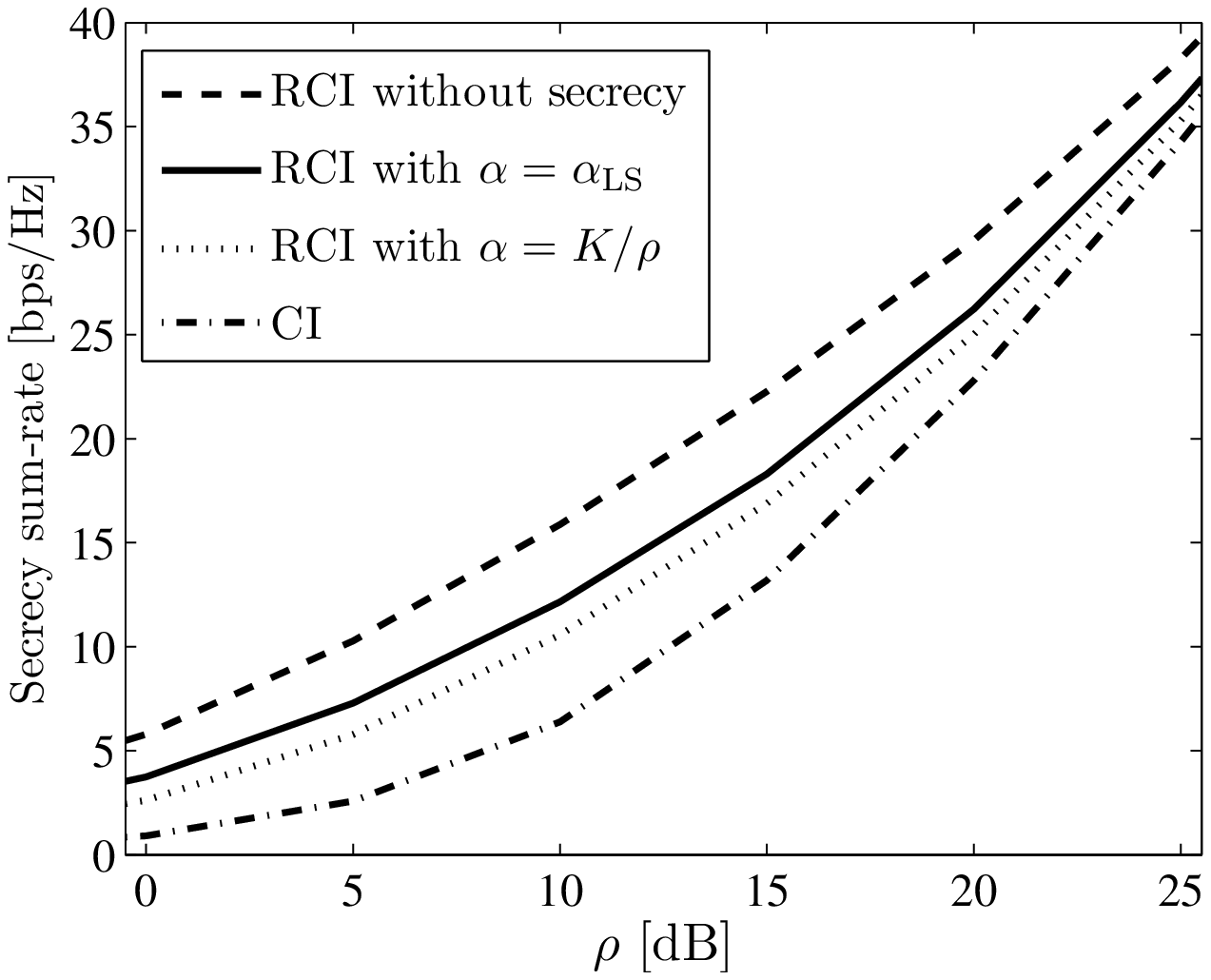}
   \label{fig:all_schemes_8}
 }
  \subfigure[$K=16$]{
   \includegraphics[width=3.1in]{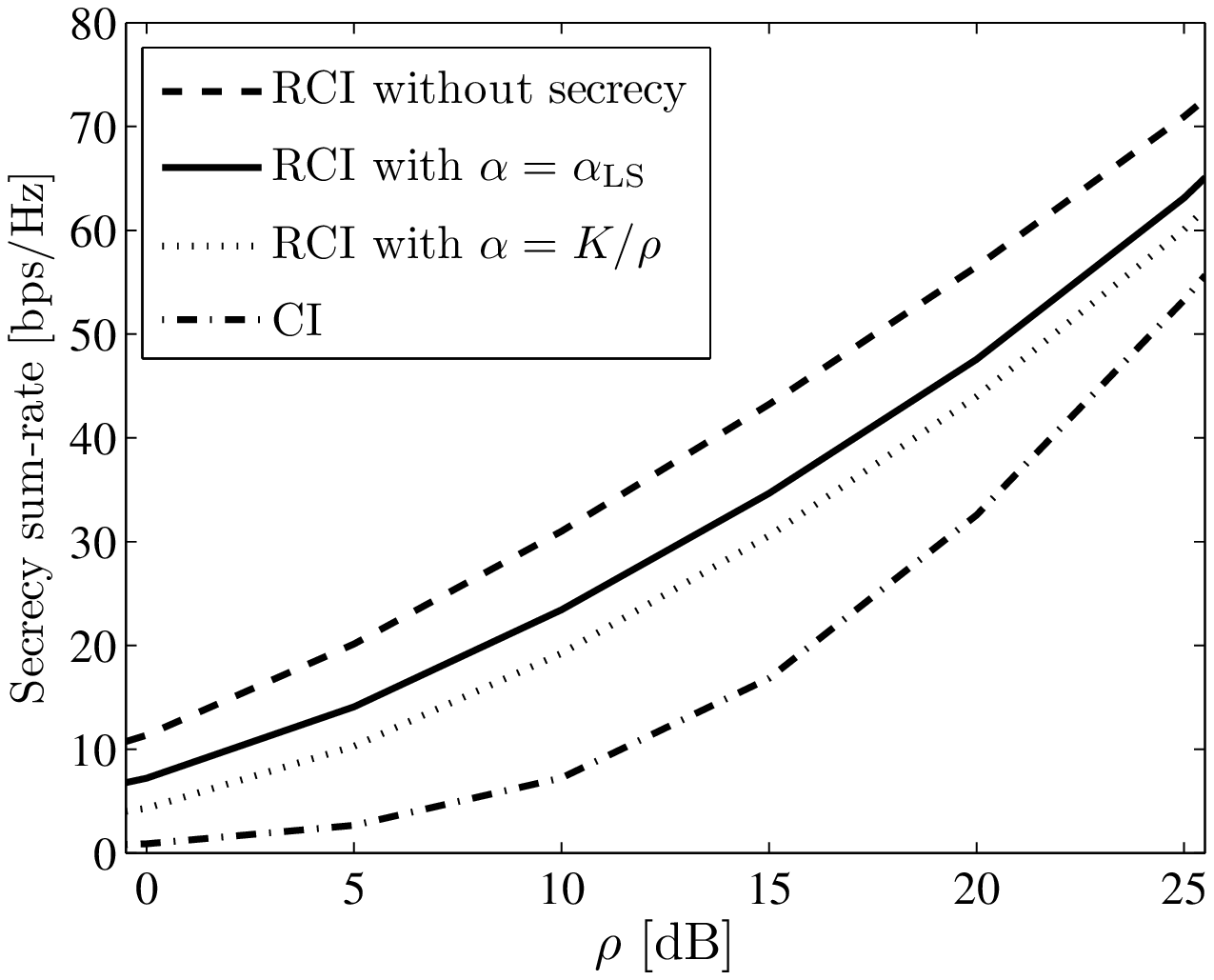}
   \label{fig:all_schemes_16}
 }
 \subfigure[$K=32$]{
   \includegraphics[width=3.1in]{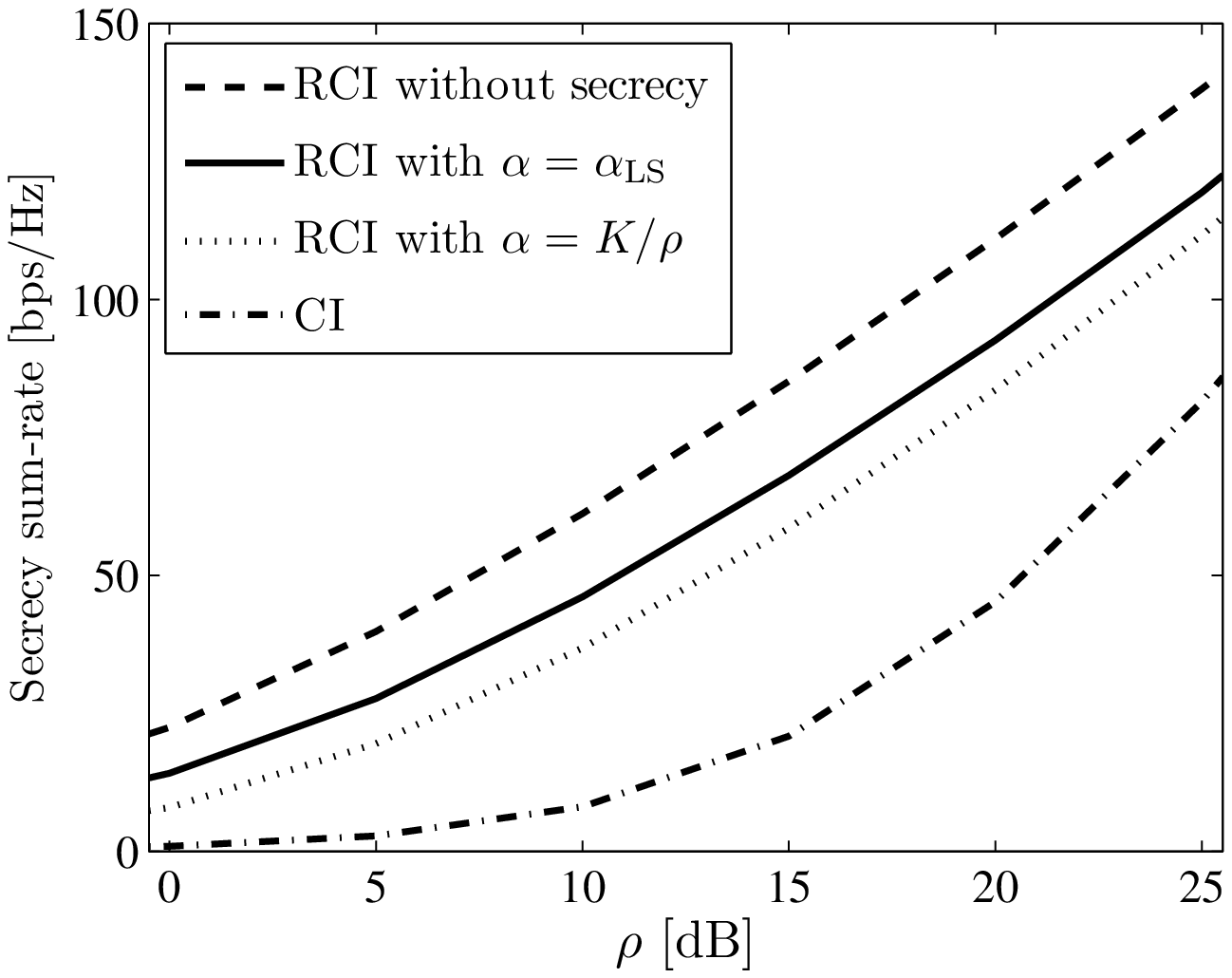}
   \label{fig:all_schemes_32}
 }
\caption{Comparison between the RCI precoder with $\alpha_{\mathrm{LS}}$ and other linear schemes. The secrecy loss is also shown as the gap between dashed and solid lines.}
\label{fig:precoding_schemes_comparison}
\end{figure}

In Fig. \ref{fig:precoding_schemes_comparison} we compare the simulated secrecy sum-rate of the RCI precoder using the large-system regularization parameter $\alpha_{\mathrm{LS}}$ with CI precoding \cite{YooJSAC06} and RCI precoding with $\alpha = K/\rho$, which maximizes the sum-rate without secrecy \cite{Peel05}. The sum-rate of the optimal RCI precoder without secrecy requirements is also plotted. The figure shows plots for $K = 4$, $8$, $16$, $32$. We observe that CI precoding exhibits a large performance loss compared to the secrecy sum-rate of the optimal RCI precoder for large values of $K$. The RCI precoder with $\alpha = K/\rho$ outperforms CI precoding, but it is suboptimal compared to the RCI precoder that uses $\alpha_{\mathrm{LS}}$. We note that although CI precoding achieves secrecy in a simple way by completely canceling the information leakage, this comes at the cost of a poor sum-rate. Secrecy can be achieved with a significantly larger sum-rate by using the RCI precoder with $\alpha_{\mathrm{LS}}$. We also observe that the secrecy loss between the sum-rate of the RCI precoder without secrecy and the secrecy sum-rate of the RCI precoder is almost constant at high SNR for large $K$. This confirms the result we derived in (\ref{eqn:secrecy_loss_large_SNR}). Moreover, the value of the simulated per-antenna secrecy loss is 0.59 bits for $K = 32$ and $\rho = 25 \mathrm{dB}$; close to the 0.62 bits suggested by the 
analysis in 
(\ref{eqn:secrecy_loss_large_SNR}).

\begin{figure}
\centering
\includegraphics[width=\columnwidth]{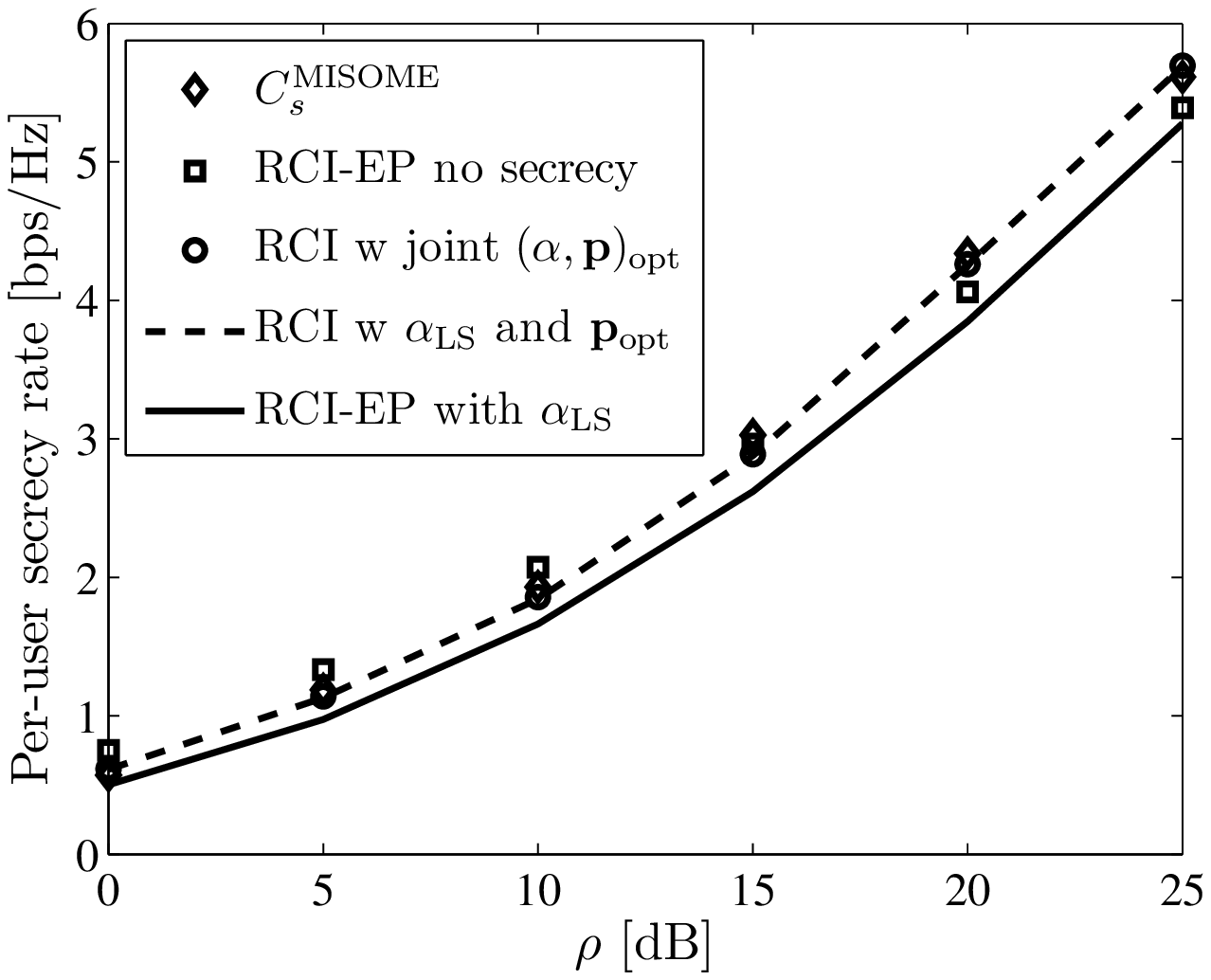}
\caption{Per-user secrecy rate vs. $\rho$ for $K=4$ users: without power allocation (solid), with $\alpha_{\mathrm{LS}}$ and $\mathbf{p}_{\mathrm{opt}}$ (dashed) and with joint optimal $(\alpha,\mathbf{p})_{\mathrm{opt}}$ (circle). The rate of the optimal RCI precoder without secrecy requirements (square) and the secrecy capacity of the MISOME channel (diamond) are also plotted.}
\label{fig:PA}
\end{figure}

In Fig. \ref{fig:PA} we compare the simulated per-user secrecy rate of the proposed precoder with jointly optimized regularization parameter $\alpha$ and power allocation vector $\mathbf{p}$ to the RCI precoder with $\alpha_{\mathrm{LS}}$ and $\mathbf{p}_{\mathrm{opt}}$, and to the RCI-EP precoder with $\alpha_{\mathrm{LS}}$. We observe that there is a negligible performance difference between the proposed precoder and the RCI precoder with $\alpha_{\mathrm{LS}}$ and $\mathbf{p}_{\mathrm{opt}}$. As a result, a low-complexity, near-optimal RCI precoder may be implemented by using $\alpha_{\mathrm{LS}}$ and optimizing the power vector separately. The figure shows that for $K = 4$, the proposed power allocation scheme always outperforms the RCI-EP precoder with $\alpha_{\mathrm{LS}}$ by up to $20$ percent, and the gain does not vanish at high SNR. This occurs because at high SNR $\xi_{\mathrm{opt}} \rightarrow 0$ and the RCI precoder behaves as a CI precoder, for which the optimal power allocation is waterfilling \cite{YooJSAC06}. Hence, equal power allocation for RCI is suboptimal at high SNR. 
Fig. \ref{fig:PA} also shows that the proposed power allocation scheme reduces the sum-rate loss due to the secrecy requirements. For $\rho \geq 15$dB, RCI with power allocation achieves a per-user secrecy rate which is even higher than the per-user rate achieved by the optimal RCI-EP without secrecy requirements. Furthermore, Fig. \ref{fig:PA} shows the simulated secrecy capacity $C_{s}^{\mathrm{MISOME}}$ of a MISOME channel with the same per-message transmitted power. Although $C_{s}^{\mathrm{MISOME}}$ is obtained in a single-user and interference-free system \cite{Khisti10I}, at high SNR, RCI with power allocation achieves a per-user secrecy rate as large as $C_{s}^{\mathrm{MISOME}}$.

%% file: Section7.tex
\section{Conclusions}

In this paper, we considered the problem of secret communication in a multi-user MIMO system with malicious users. We proposed a linear precoder based on regularized channel inversion (RCI) with a regularization parameter and power allocation vector that maximize the achievable secrecy sum-rate. The analysis presented in the paper, as well as the simulation results, showed that RCI with equal power allocation (RCI-EP) and with the optimal regularization parameter outperforms several other linear precoding schemes. Moreover, it achieves a sum-rate that has same scaling factor as the sum-rate of the optimum RCI precoder without secrecy requirements. The secrecy requirements result in a loss in terms of the sum-rate. This loss can be compensated by the proposed power allocation scheme, which increases the secrecy sum-rate compared to RCI-EP.

Part of the analysis presented in this paper focused on the case when the number of users $K$ equals the number of transmit antennas $M$. Generalizing the results to the case when $K$ and $M$ can take any value is part of our ongoing research effort. When $K>M$, the secrecy sum-rate degrades due to the increased interference and information leakage. Therefore, it can be useful to design a user scheduling algorithm that selects a subset of the users to be served, thus increasing their SINR. However, user scheduling cannot reduce the number of potentially malicious receivers in the network, since discarded users are still able to eavesdrop. The transmission of artificial noise can limit the eavesdropping ability of the discarded users, but it must be harmless to intended receivers. 

Throughout the paper, we focused on the worst-case scenario when the transmitter assumes that users cooperate and jointly eavesdrop on other users. We are interested in this scenario because the transmitter is unable to predict whether the users are eavesdropping or not. Possible extensions of this work include considering a scenario where only some of the users eavesdrop on other users, or where users can individually eavesdrop, but without cooperating. In this case, the secrecy rate for each user is limited by the eavesdropper that receives the largest information leakage. We leave the analysis of these aspects for future work.
\vspace*{-0.1cm}